\documentstyle[11pt,newpasp]{article}

\begin{document}

\title{On the Formation of Boxy and Disky Elliptical Galaxies}
\author{Andreas Burkert and Thorsten Naab}
\affil{Max-Planck-Institut for Astronomy, K\"onigstuhl 17, D-69117 Heidelberg,
Germany}

\begin{abstract}
N-body simulations of major spiral-spiral mergers with different mass ratios
$q=M_1:M_2 \leq 4:1$ are presented. All mergers lead to
spheroidal stellar systems with $r^{1/4}$-like surface density profiles, resembling
elliptical galaxies. The internal dynamical structure of the remnants does
however depend critically on the adopted mass ratio. Equal-mass mergers with
$q \leq 2:1$ form radially anisotropic, slowly rotating systems with
preferentially boxy isophotes. Unequal-mass mergers with $q > 2:1$ result in
the formation of rotationally supported ellipticals with disky isophotes and
small minor axis rotation. Projection effects lead in general
to a large scatter in the kinematical and isophotal properties.
If the observed dichotomy between boxy and disky ellipticals is a result of
different mass ratios of the merger components massive
ellipticals must have formed preferentially from equal-mass 
mergers in contrast to low-mass ellipticals which had to form from 
unequal-mass mergers. 
\end{abstract}


\section{Introduction}

Elliptical galaxies can be subdivided into two classes with distinct kinematical and
orbital properties. Whereas low-luminosity ellipticals are in general isotropic and
flattened by rotation ($\log (v_m/\sigma)^* \approx 0$),
high-luminosity ellipticals are anisotropic, slowly rotating
with ellipticities that result from anisotropy in their velocity dispersion
($log (v_m/\sigma)^* \approx -1$).

Bender (1988) and Bender, D\"obereiner \& M\"ollenhoff (1988; 
see also Kormendy \& Bender 1996 and references therein) discovered deviations of the
isophotal shapes of ellipticals from perfect ellipses that correlate with other 
kinematical properties: massive anisotropic ellipticals show boxy
isophotes with a fourth order Fourier coefficient $a_4 < 0$ 
whereas low-mass, rotationally supported ellipticals are characterized by
disky deviations from perfect ellipses with $a_4 > 0$. 
It is reasonable to assume that the different
isophotal shapes reflect different internal orbital distributions: 
boxy ellipticals are dominated by stars on box-orbits, disky ellipticals
contain a dominant stellar population on tube-orbits. 
Various correlations exist
between the observed isophotal and kinematical properties 
of ellipticals. For example, for disky ellipticals the $a_4$-coefficient decreases
with increasing ellipticity e whereas for boxy ellipticals $a_4$ increases with
increasing e. In addition, no elliptical has been found which is strongly
anisotropic but disky although
there exists no theoretical reason why such an equilibrium system should not
exist.

Toomre \& Toomre (1972) proposed that early type galaxies originate from major mergers
of disk galaxies. This merger hypothesis has been tested and investigated in great
details by numerous authors using numerical simulations (see Barnes \& Hernquist 1992
for a review). 
Steinmetz \& Buchner (1995) noted departures from pure ellipses in equal-mass merger
remnants and Barnes (1992) and Heyl et al. (1996) found misalignments between the
spin axis and the minor axis in major merger remnants. Barnes (1998) investigated
unequal-mass mergers and found that the merger remnants show disky morphology when
viewed edge-on. 

\section{The Merger Models}

We have performed very high-resolution N-body simulations
of spiral-spiral mergers with different mass ratios q. The spirals are constructed in
dynamical equilibrium using the method described by Hernquist (1993). Each galaxy
consists of an exponential disk, a spherical, non-rotating bulge and a dark halo
component. As the isophotal and 
kinematical structure of ellipticals various with radius it is a matter of debate
how to define global kinematical properties.
In order to compare our simulations with observations
we follow as closely as possible the analysis and definition of Bender et al. (1988)
by generating an artificial image of the remnant which is smoothed with a Gaussian filter.
The isophotes and their deviations from perfect ellipses are then determined using
a reduction package kindly provided by Ralf Bender. The method as well as details about
the initial conditions and the numerical method are described in 
Naab, Burkert \& Hernquist (1999).

\section{Analysis of the merger remnants}

The mean radial distribution of $a_4$ along the major axis does in general
vary with radius.
There is however a clear trend for equal-mass mergers to have boxy isophotes with
$a_4$ becoming more negative outside of 0.4 r$_e$, where $r_e$ is the effective
radius of the system. Unequal-mass mergers have disky deviations
with positive $a_4$-values inside $r_e$ with a peak
at $r=0.5 r_e$. 

Figure 1 shows the characteristic isophotal and kinematical properties of a
1:1 merger (filled circles) and of a 3:1 merger (open circles) for 200 random 
projections. Note that the large spread is a result of projection effects.
One can however clearly see that the equal-mass merger produces remnants that
are anisotropic and boxy with large minor axis rotation whereas the 3:1 merger
forms an isotropic and disky elliptical with small minor axis rotation.
In addition, the diskiness increases with increasing ellipticity for
the 3:1 merger whereas $a_4$ becomes more negative with increasing ellipticity
for the boxy merger remnant (1:1 merger).
These results are in excellent agreement with the observations.

\begin{figure*}
\includegraphics{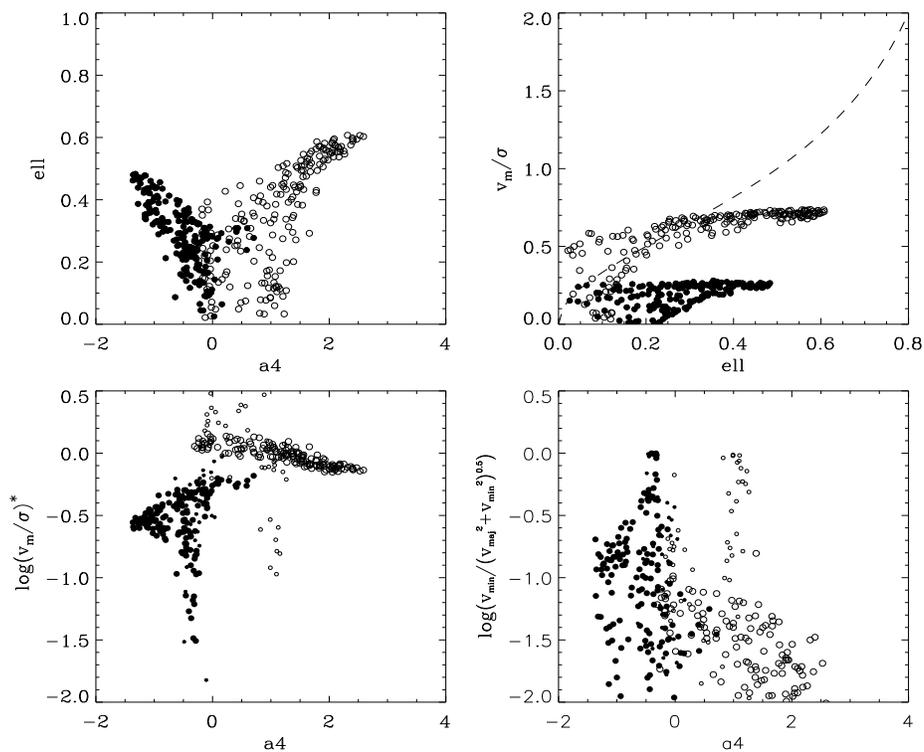}
\vspace{9.5cm}
\caption{Kinematical and photometric properties for 200 random 
projections of the merger remnants.
Filled circles show the values for the equal-mass merger
while open circles show the values for the 3:1 merger.
{\it Upper left panel}: Ellipticity versus $a_4$.
{\it Upper right panel}: Rotational velocity over central velocity dispersion
versus ellipticity. {\it Lower left panel}: Anisotropy parameter versus $a_4$.
{\it Lower right panel}: Minor axis rotation versus $a_4$,
with $v_{maj}$ and $v_{min}$ being the maximum velocity along the
major and minor axis, respectively.}
\end{figure*}

\section{Discussion}
The previous section showed the results of a 1:1 and a 3:1 merger. 
Additional simulations with different mass ratios and different orbital 
parameters show that mergers with mass ratios $q < 2$ always form boxy ellipticals 
whereas mergers with q $>$ 2 lead to disky ellipticals.
The observed dichotomy between boxy and disky ellipticals could therefore 
originate from variations in the mass ratios of the merger
components. 

Projection effects do not change the fundamental difference between equal
and unequal-mass merger remnants. They do however lead to a large spread in the
global parameters and to trends between the ellipticity and the $a_4$-coefficient
that are in agreement with the observations. 

Observations show that disky  ellipticals have on average lower luminosities than
boxy ellipticals. Our results would then indicate that low-mass ellipticals
formed preferentially from unequal-mass mergers whereas equal-mass
mergers dominate the formation of high-mass ellipticals. This result is 
puzzling as there does not exist a convincing cosmological argument for such a scenario.

\acknowledgments

We thank Ralf Bender, Hans-Walter Rix for helpful discussions and
Lars Hernquist for a copy of his program to generate the initial
conditions for spiral-spiral mergers.

\end{document}